\documentstyle[prl,floats,aps,epsfig]{revtex}

\begin{document}
\draft
\twocolumn[\hsize\textwidth\columnwidth\hsize\csname %
@twocolumnfalse\endcsname

\title{Phase transitions in the mesoscopic superconducting square}
\author{$^{a,b}$J.Bon\v ca and $^a$V.V. Kabanov}
\address{$^a$J. Stefan Institute 1001, Ljubljana, and
$^b$FMF, University of Ljubljana, Ljubljana, Slovenia}
\date{\today}
\maketitle

\begin{abstract}\widetext
We solve the Ginzburg-Landau equation (GLE) for the mesoscopic thin
film of the square shape in the magnetic field. In the limit of
Ginzburg-Landau parameter $\kappa \to \infty$ we find a series of
first and second order phase transitions as temperature and/or
magnetic field changes.  First order phase transitions between giant
flux states can be described with a simple variational procedure.  We
discuss the similarity with rotating liquid $He^{4}$ and derive a
simple formula for $H_{c1}$. We identify order parameters based on
symmetry arguments and we propose Landau functional describing the
second order phase transition.
\end{abstract}
]

\narrowtext

Advances in nanotechnology and constantly shrinking semiconductor
devices have motivated researches to study properties of mesoscopic
superconducting samples. One line of research in this field has
focused on the problem of the phase transitions in the mesoscopic
superconducting sample under the influence of the external magnetic
field \cite{mosch}.  There are two characteristic limits in which phase
transitions have different properties. If the size of the sample $a\gg
\xi $, where $\xi $ is the superconducting coherence length, and
applied field are large enough, there are many vortices in the
sample. In this case long-range interaction between vortices and image
vortices is screened by spontaneous creation of vortex loops near the
sample boundary\cite{kus}. It leads to the decrease of the surface barrier for
the vortex to penetrate into the sample. In the opposite case, when $a\sim \xi
\ll \lambda $, with $\lambda$ being the London penetration depth, there are
only few vortices in the sample. The standard Abrikosov approach \cite{abr}
must be modified because of the strong influence of the sample boundaries. In
this case magnetic field is almost uniform throughout the sample and
long-range forces between vortices are not important. Thermodynamics of this
system is determined by the short-range repulsion of vortices and
Bean-Livingston barrier forces\cite{bean}.

Different approaches have been applied for the investigation of
phase transitions in the latter limit. Most of them consider disk
geometry. Buzdin and Brison applied electrostatic formalism to
consider influence of the barrier on the vortex structure of the thin
superconducting disk \cite{buz}. Within this approach vortices are
replaced by the hard-core particles interacting through Coulomb forces
and the giant vortex has never been discussed. Numerical solution of
GLE for the same geometry reveals
a series of the first and second order phase transitions in the
superconducting disk. Such  transitions take place between giant
vortex states with different vorticity as well as between a giant vortex
state and a multi-vortex state as the external field changes
\cite{Deo,Schw,Peet}. We emphasize an important
difference between the disk and the square geometry. Solution of the
linearized GLE, describing the nucleation of superconducting order
parameter near $H_{c2}$ line for the disk, always corresponds to the giant
vortex state. On the other hand, as it was demonstrated by Chibotaru
{\it et al.} \cite{mosch}, there are many  well separated zeros of the
order parameter in the case of the square sample. Consequently, the  behavior
of the square sample near $H_{c2}$ line should be qualitatively different
from the disks. On the basis of the solution of the linearized GLE the
appearance of the antivortex in the center of the sample
has been predicted\cite{mosch}.

In the this paper we investigate phase transitions in the
superconducting film of a square shape as a function of temperature
$T$ and external magnetic field $H$ in the limit $a\sim \xi \ll
\lambda $. We solve GLE for the thin superconducting film with the
thickness $d\ll \xi $. We also assume that superconducting film is an
extreme type two superconductor, where the Ginzburg-Landau parameter
$ \kappa =\lambda /\xi \gg 1$. We
show that configuration with one anti-vortex in the center and four
vortices on the diagonals of the square is unstable when we move away
from the $H_{c2}$ line and nonlinear term in the GLE is considered. On
the contrary, at higher magnetic field, the configuration with four
vortices on diagonals of the square remains stable. We find a sequence
of phase transitions of the first order between giant vortex states as
well as between multi-vortex states with different vorticity.  Second
order phase transition takes place when a giant vortex state splits
into a multi-vortex state with simultaneously breaking the $C_4$
symmetry.  Such transitions are discussed in terms of the phenomenological
theory of Landau.

GLE for dimensionless complex order parameter $\psi$  has the following form:
\begin{equation}
\xi^{2}(i\nabla +{\frac{2\pi {\bf A}}{{\Phi_{0}}}})^{2}\psi - \psi +\psi
\vert \psi \vert ^{2}=0
\end{equation}
here $\xi = {\frac{\hbar^{2}}{{4m\vert \alpha \vert}}}$, $\alpha$ is
temperature dependent parameter of Ginzburg-Landau expansion for the free
energy, $\Phi_{0}$ is the flux quantum, ${\bf A}$ is the vector potential $%
{\bf H}=\nabla\times {\bf A}$. 
The second GLE equation for the vector potential can be written as:
\begin{equation}
\nabla \times \nabla \times {\bf A} = -i{\frac{\Phi_{0}}{{4\pi \lambda^{2}}}}
(\psi^{\ast}\nabla\psi-\psi\nabla\psi^{\ast}) -{\frac{\vert\psi\vert^{2}{\bf %
A} }{{\lambda^{2}}}}.
\end{equation}
Since we consider the case of a small mesoscopic square where
$a \sim \xi \ll \lambda$, the magnetic field is uniform in the film.
The correction to the
external field is  of the order of $1/\kappa^{2}$ and may be found by 
solving Eq.(1) while assuming uniform magnetic field and substituting
the solution of Eq.(1) to Eq.(2).
Such a solution is equivalent to the expansion of the free
energy in  $1/\kappa^{2}$ series. In addition to Eq.(1) we have to supply the
boundary condition for the superconductor-insulator junction:
\begin{equation}
(i\nabla + {\frac{2\pi {\bf A}}{{\Phi_{0}}}}){\bf n}\psi = 0,
\end{equation}
where $\bf n$ is normal vector to the surface of the sample.

Introducing $N\times N$ discrete points in the square we rewrite Eq.(1) in
the form of nonlinear discrete Schr\" odinger equation:
\begin{equation}
\sum_{{\bf l }}t_{\bf i+l ,i}\psi_{\bf i+l }-4t_{\bf i,i}%
\psi_{\bf i}-\psi_{\bf i}+\psi_{\bf i}|\psi_{\bf i}|^{2}=0
\end{equation}
where summation index ${\bf l}=(\pm 1,0)$, $(0,\pm 1)$ points toward
nearest neighbours and $t_{\bf i_{1},i}=(\xi N/a)^{2}
exp(-{\frac{2\pi i }{{\Phi _{0}}}}\int_{{\bf i}}^{{\bf i_{1}}}{\bf A}
({\bf r})d{\bf r})$\cite{km}. Equivalent
discretization of the boundary conditions, Eq.(3), provides an additional
equation which can be directly solved and substituted into Eq.(4). As a result,
equation close to the boundary are slightly different from the 'bulk':
\begin{equation}
\sum_{{\bf l }}t_{\bf i+l ,i}\psi_{\bf i+l }-\epsilon (%
{\bf i})t_{\bf i,i}\psi_{\bf i}-\psi_{\bf i}+\psi_{\bf i}|\psi_{\bf i}|^{2}=0,
\end{equation}
where $\psi_{\bf i}=0$ if ${\bf i}$ is outside of the sample, $\epsilon (
{\bf i})=4-\delta _{i_{x},1}-\delta _{i_{x},N}-\delta _{i_{y},1}-\delta
_{i_{y},N}$ and ${\bf i}=(i_{x}=1,\dots,N,i_{y}=1,\dots,N)$.
There is one important advantage of such a treatment of the
boundary condition. When neglecting the nonlinear term in the Eq.(4), the
system of linear equations reduces to the problem of eigenvalues and
eigenfunction of the $hermitean$ matrix.
On the other hand, the solution of nonlinear
equations requires iterations and inversion of the hermitean matrix.

Let us first discuss the solution of the linearized GLE and compare
our results with previous studies \cite{mosch}.  The lowest eigenvalue
of the linear GLE determines the upper critical field of the
sample. We have calculated eigenvalues of the linear problem expressed
in units $(a/\xi(T))^{2}$ as a function of the dimensionless external
magnetic field $h=\Phi/\Phi_{0}$ where $\Phi$ is the total flux
through the sample. Our results for a few lowest eigenvalues agree
within the linewidth with the results of Ref.\cite{mosch}. Spatial
pattern of the order parameter is also similar. For the field $h\simeq
5.5$ we have observed 5 zeros of the order parameter near the center
of the square. The solution corresponds to 4 vortices on the diagonals
and one anti-vortex in the center of the square with total vorticity
$m=3$\cite{mosch}. The distance between vortices is of the order of
$\delta \simeq 0.12\xi << \xi$.  All these zeros are pinned in the
region of the square where $|\psi(x,y)|$ is small (four orders less
then the value of the order parameter near the sample boundary). This
indicates that vortex-anti-vortex structure becomes unstable when we
move away from the $H_{c2}$ line and nonlinear term and $1/\kappa$
corrections are considered. Moreover, the value of the order parameter
and the screening current are small near zeros. In that case
correction to the external field is determined by the current flowing
around all 5 zeros of the order parameter rather than by the current
between them. Therefor, we do not expect suppression of the field in
the core of the anti-vortex.

\begin{figure}[tb]
\begin{center}
\epsfig{file=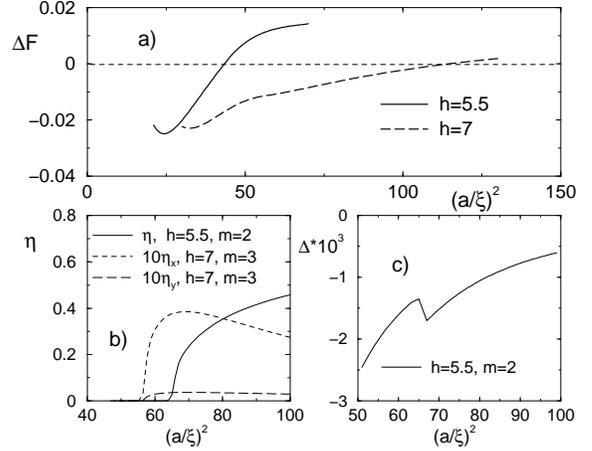,height=75mm,angle=-90}
\end{center}
\caption{ a) Difference in the free energy $\Delta F$ in units of
$\frac {\Phi_{0}^{2}a^{2}d}{(4\pi)^{3}\lambda^{2}\xi^{2}}$ between
solutions with different integral vorticities. For the case of $h=
5.5$, $\Delta F = F(m=3)-F(m=2)$ and for $h= 7$,
$\Delta F = F(m=4)-F(m=3)$. b) Order parameter
$\eta$ vs. $(a/\xi )^{2}$ for the solution with $m=2$
calculated at $h=5.5$ and two dimensional order parameter $\eta_x$ and
$\eta_y$ for the solution with $m=3$ calculated at $h=7$. c)
The second derivative of the free energy $\Delta = d^2F/d((a/\xi
)^{2})^2$ for the solution with $m=2$ at $h=5.5$. }
\label{f5.5}
\end{figure}

In the following we consider the changes in the vortex structure when we
move away from the $H_{c2}$ line. 
%
Taking into account the  nonlinear term, we were
unable to detect more than one zero of the order parameter near the
origin of the square for the value of the field $h=5.5$.
Consequently, the  solution with more than one zero survives only very
close to $H_{c2}$ line. We do not expect any phase transition at the
point where all zeros are joined together since total flux and the
symmetry of the solution do not change. This situation is 
different from the case of higher field $h=7$, where
solution with four zeros of the order parameter survives far from the
$H_{c2}$ line.  In fact, the giant flux solution was not detected in
that case.

\begin{figure}[]
\begin{center}
\epsfig{file=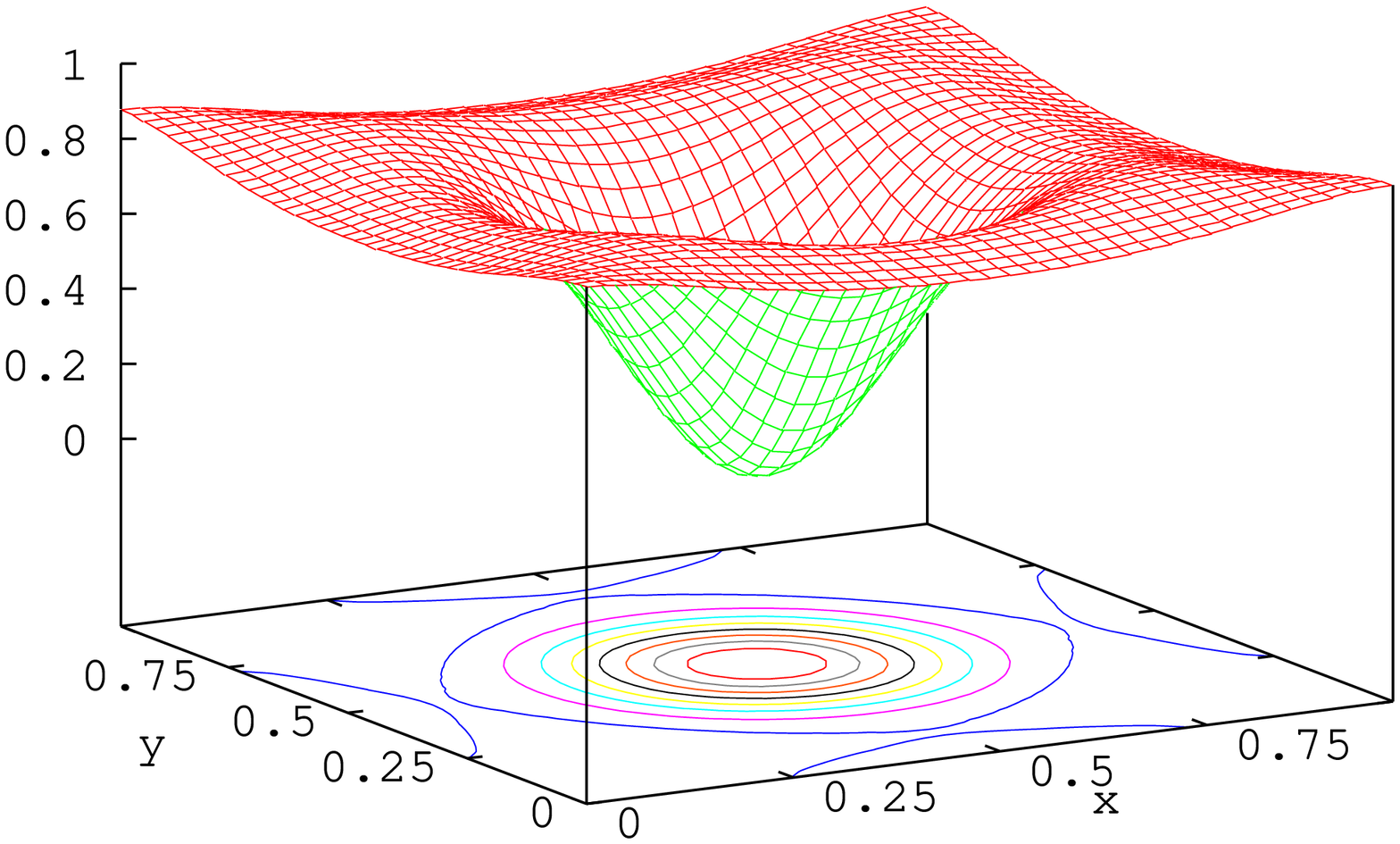,height=75mm,angle=-90}
\epsfig{file=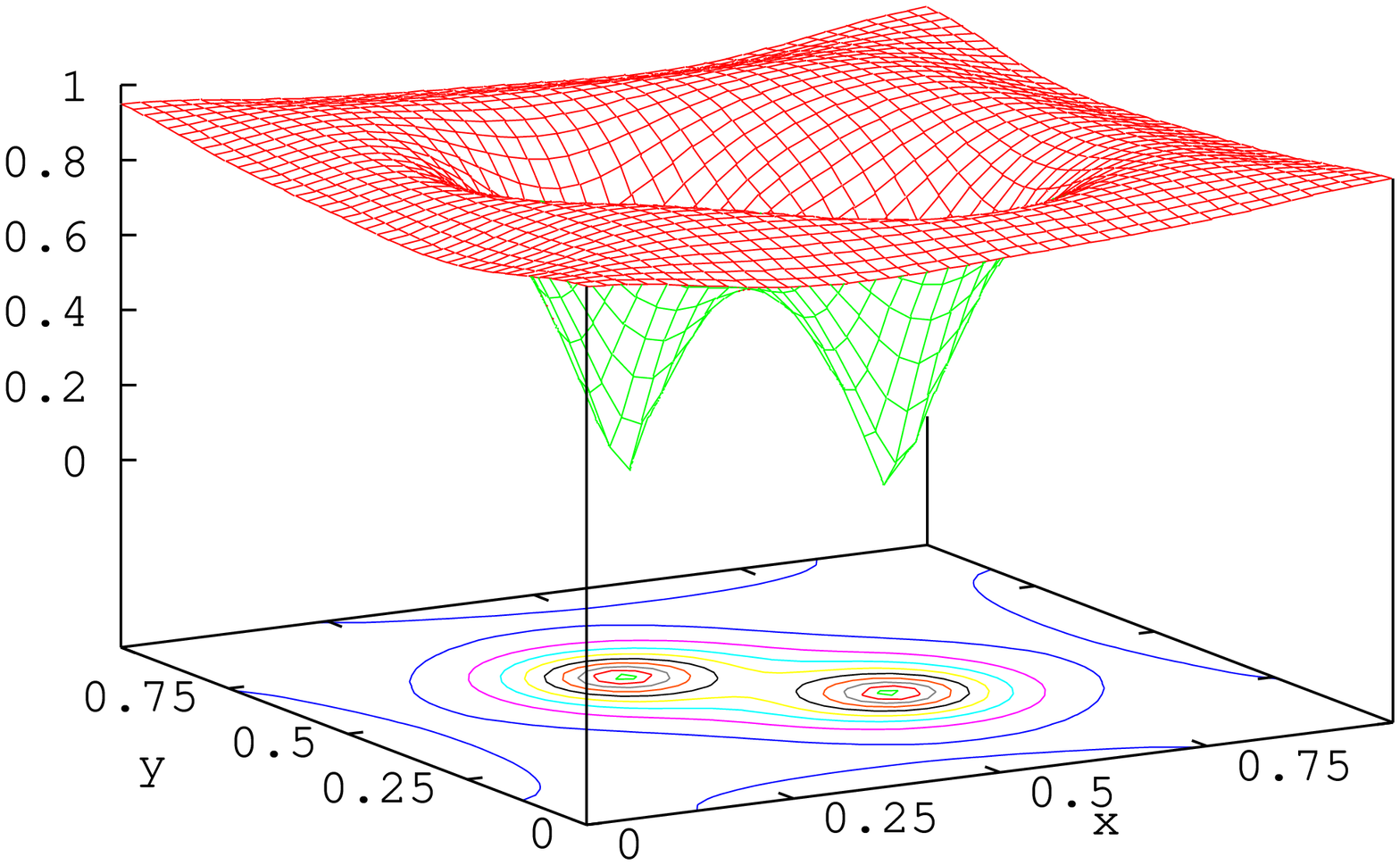,height=75mm,angle=-90}
\end{center}
\caption{$|\psi(x,y)|$ calculated at fixed magnetic field $h=
5.5$. Presented are solutions with lowest free energy calculated at
$(a/\xi)^2=50$ with one giant flux $m=2$ (top) and
$(a/\xi)^2=90$ with two separated fluxes each carrying $m=1$
(bottom). Contours represent $|\psi(x,y)|=0.1,\dots,0.9$.}
\label{psi1}
\end{figure}

In the Fig.~(\ref{f5.5}a) we plot the difference in the free energy
between solutions with different integral vorticity $\Delta F =
F(m=3)-F(m=2)$ as a function of $(a/\xi )^{2}$ for the fixed magnetic
field $h= 5.5$.  As it is clearly seen from the figure, near $(a/\xi
)^{2}\sim 43$ a first order phase transition takes place. At that
point the high-temperature phase corresponding to the giant vortex
with $m =3$ becomes metastable, and the phase corresponding to giant
vortex, shown in Fig.~(\ref{psi1}~top), with $m=2$ becomes a ground
state.  At this point, the slope of the first derivative of the free
energy as a function of $T$ is discontinuous, which corresponds to the
latent heat of the transition. With further decrease of the
temperature and $\xi$, the second transition takes place.  At
$(a/\xi)^{2}\sim 66$ the giant vortex located in the center splits
along one of the diagonals, Fig.~(\ref{psi1}~bottom). This transition
is the second order phase transition, where $|\psi|^{2}$ is no longer
invariant under four-fold axis of the square. The phase transition is
clearly observed by computing the order parameter $\eta =\int xy|\psi
(x,y)|^{2}dxdy$, presented in Fig.~(\ref{f5.5}b). A nonvanishing
$\eta$ is followed by a jump in the second derivative of the free
energy $\Delta = d^2F/d((a/\xi )^{2})^2$, presented in
Fig.~(\ref{f5.5}c). We estimated the magnitude of the effect in terms
of the specific heat jump: $\Delta C /a^{2}d = \frac {5\cdot 10^{-3}
\Phi_{0}^{2}T_{c}^{'}} {(4\pi)^{3}\lambda^{2} \xi^{2} T_{c}^{2}}$,
where $T_{c}^{'}$ is the critical temperature of the transition
determined by the condition $(a/\xi)^{2} \simeq 66$, and $T_{c}$ is
the critical temperature of the sample at $h=0$.  Landau functional,
describing this phase transition, is defined as ${\em F}=\alpha_1 \eta
^{2}+\beta_1 \eta^{4}$, where $\alpha_1 $ and $\beta_1 $ are Landau
coefficients with $\alpha_1 \propto (T-T_{c}^{'})$. This phase
transition corresponds to the one dimensional corepresentation $B$ of
the nonunitary $C_{4v}(C_{4})$ group.

Since we restrict our calculation to the limit of $\kappa \to \infty$
and $\lambda \gg a$, long range forces ($r \gg \xi$) between vortices
are irrelevant. At short distances $r \sim \xi$ there is short range
repulsion between them. There is another interaction of vortices with
the boundaries of the sample, known as Bean-Livingston barrier
\cite{bean}. In the vicinity of $H_{c2}$ line the interaction with the
boundaries is larger than the repulsion between vortices. This is the
case for the magnetic field $h=5.5$, where the giant vortex with $m
=3$ is located in the center of the sample.  With the decrease of
$\xi$ the energy difference between two different solutions with
different vorticity decreases and the first order phase transition to
a giant flux state with vorticity $m=2$ takes place at $(a/\xi)^{2}
\simeq 43$. A giant flux state (with lower vorticity) remains stable,
since the interaction with the sample boundaries still dominates over
the repulsion between vortices. Further decrease of $\xi$ leads to the
decrease of the interaction of the vortices with the
boundaries. Bean-Livingston force decreases and repulsion of the
vortices in the giant flux state starts to dominate. As a result of
the interplay of the vortex-vortex repulsion and the repulsion of
vortex from the boundaries, the second order phase transition (at
$(a/\xi)^{2} \simeq 66$) between the giant vortex state and the
multivortex state takes place, preserving the integral vorticity.
Separation between vortices is determined by the vortex-vortex
repulsion that tends to separate them as far as possible, while in
the contrary, repulsion from the boundaries prevents vortices from
approaching the boundaries.

\begin{figure}[]
\begin{center}
\epsfig{file=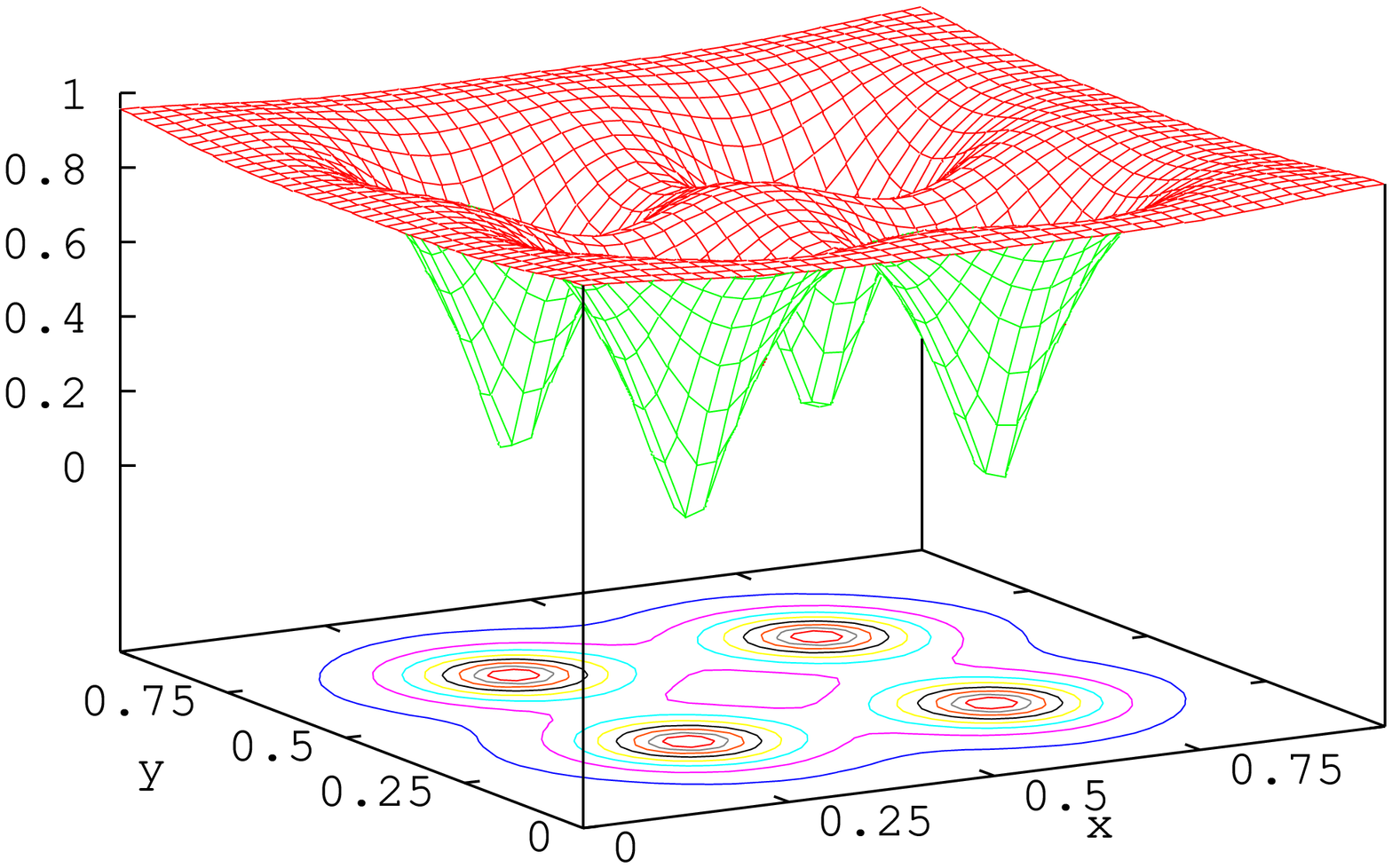,height=75mm,angle=-90}
\epsfig{file=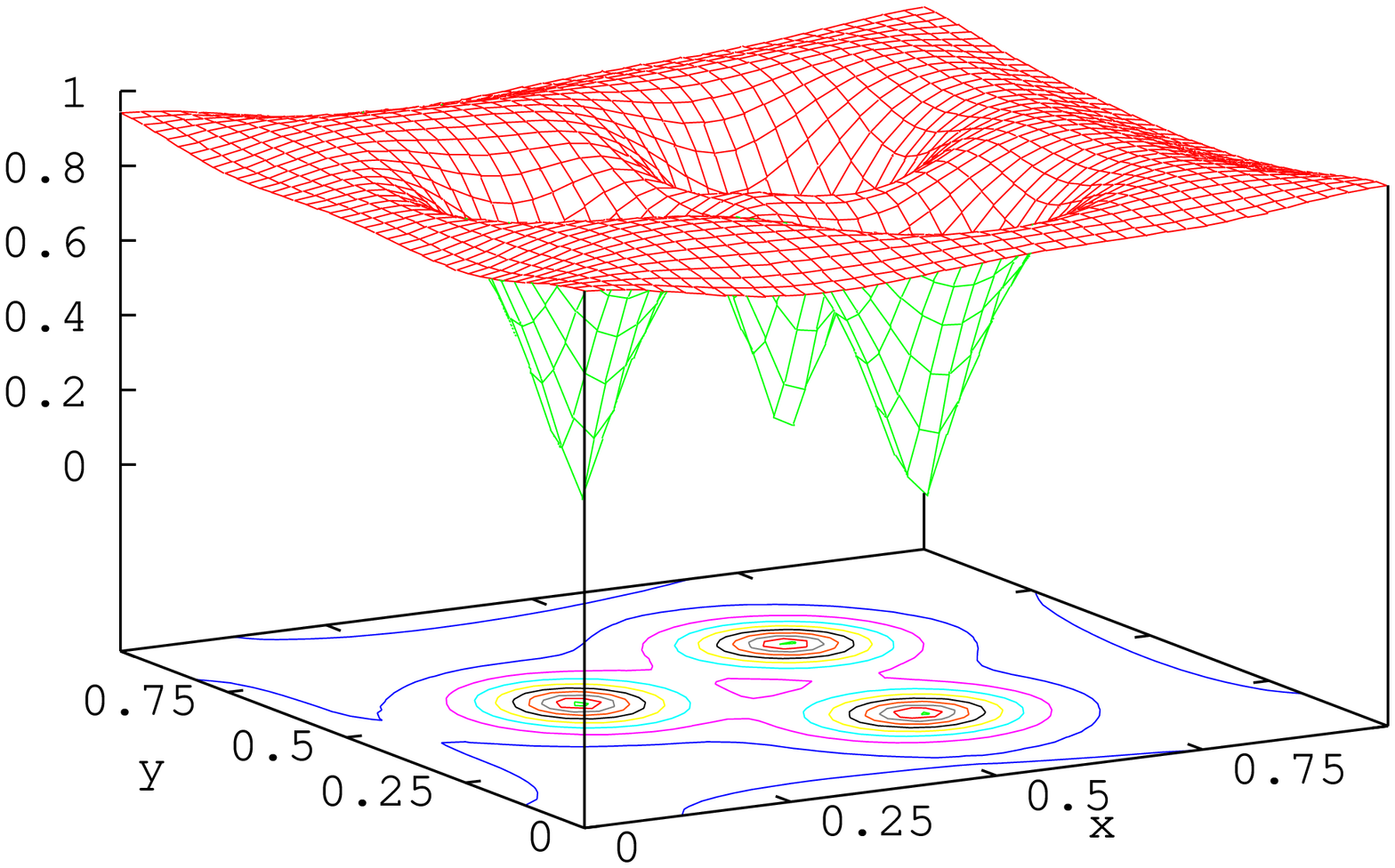,height=75mm,angle=-90}
\end{center}
\caption{$|\psi(x,y)|$ calculated at fixed magnetic field $h=
7$. Presented are solutions with lowest free energy calculated at
$(a/\xi)^2=100$
(top) and
$(a/\xi)^2=120$ 
(bottom). Contours are defined as in Fig.~(\ref{psi1}).}
\label{psi2}
\end{figure}

Situation is different when external magnetic field is increased to
$h\simeq 7$. In that case near $H_{c2}$ line ground state corresponds
to a multivortex state with the total vorticity $m =4$,
Fig.~(\ref{psi2}~top). When temperature decreases, the first order phase
transition takes place at $(a/\xi )^{2}=110$, see Fig.~(\ref{f5.5}a).
At this point
multivortex state with $m =4$ becomes unstable while  
the multivortex state with $m =3$, presented in 
Fig.~(\ref{psi2}~bottom), represents the solution with the lowest free
energy.  Apart from the change of the vorticity, the 
symmetry is also reduced at the transition  point. Phase transition takes
place in accordance with joint corepresentation $E^{\pm}$ of the
nonunitary group $C_{4v}(C_{4})$. Consequently, four
orientations of the pseudodipolar moment of the vortices are possible.
Two component order parameter corresponding to given change of
symmetry, presented in Fig.~(\ref{f5.5}b), can be determined as
follows: $\eta_{x} =\int x|\psi (x,y)|^{2}dxdy$, $\eta_{y} =\int
y|\psi(x,y)|^{2}dxdy$.  Free energy in that case depends on the
vorticity $m$ and order parameter $\eta_{x},\eta_{y}$.  ${\em
F}(m=4,\eta_{x},\eta_{y})$ always has a minimum at $\eta_{x}=
\eta_{y}=0$. For $m=3$, ${\em F}(m=3,\eta_{x},\eta_{y})=\alpha_2
(\eta_{x}^{2}+\eta_{y}^{2}) + \beta_2 (\eta_{x}^{4}+ \eta_{y}^{4}) +
\gamma_2 \eta_{x}^{2}\eta_{y}^{2}$, where $\alpha_2, \beta_2,
\gamma_2$ are Landau coefficients, $\alpha_2 \propto (T-T_{c}^{'})$
where $T_{c}^{'}$ is the temperature of the transition between the
giant vortex state and the multivortex state for the case of $m
=3$. $T_{c}^{'}$ is determined by the condition $(a/\xi)^{2} \simeq
56$. This transition is unobservable because $T_{c}^{'}$ is lower than
the transition temperature for the first order phase transition where
the vorticity $m$ changes form $m=3$ to $m=2$ (compare
Figs.~(\ref{f5.5}a) and ~(\ref{f5.5}b)).

The first order phase transition between giant vortex states with
different vorticity can be described qualitatively on the basis of
a simple variational function for the order parameter. Spatial
dependence of the order parameter in the giant vortex state with
vorticity $m$ can be approximated by the function:
\begin{equation}
\psi(r,\phi) = \left \{ \begin{array}{r}
(r/\xi)^{m}\exp{(im\phi)};\hspace{0.5cm} r<\xi,\\
\nonumber
\exp{(im\phi)};\hspace{0.5cm} r>\xi.
\end{array} \right . 
\end{equation}
Substituting this function into the Ginzburg-Landau functional and
keeping only leading terms in $(a/\xi)^{2}$, we obtain a simple expression
for the free energy which should be minimized as a function of
vorticity $m$:
\begin{equation}
F\simeq \frac {\Phi_{0}^{2}d}{(8\pi)^{2}\lambda^{2}}
(m^{2}\ln{(a^{2}/\pi \xi^{2})}/2- \Phi \cdot m/\Phi_{0}).
\end{equation}
Minimization of the Eq.(7) in the large $m \gg 1$ limit provides the
expression for the vorticity:
\begin{equation}
m \simeq \frac{\Phi}{\Phi_{0}}/\ln{(a^{2}/\pi \xi^{2})}.
\end{equation}
Minimization of the Eq.(7) for $h=5.5$ provides that the phase transition
from $m=3$ to $m=2$ takes place at
$(a/\xi)^{2}=28$ which is lower than calculated value $(a/\xi)^{2}=43$.
Nevertheless, this simple variational formula provides simple
explanation of the transition with the change of vorticity in the
giant vortex. It is interesting to note that due to logarithmic term,
Eq.(7) for the free energy is similar to
the free energy of the rotating superfluid liquid \cite{halat}. This
similarity appears because $\lambda > a$ and all integrals are
cut at $a$, rather than at $\lambda$. Eq.(7) yields  also the  estimate
for $H_{c1}$,
the field at which the first vortex appears in the sample. Substituting
$m=1$ to the Eq.(1) and solving equation $F(m=1)=0$ we obtain:
\begin{equation}
H_{c1}=\frac {\Phi_{0}}{2 a^{2}} \cdot \ln{(a^{2}/\pi \xi^{2})}.
\end{equation}
This expression is similar to the bulk $H_{c1}$ where $\lambda^{2}$ is
substituted to $a^{2}/\pi$.

In conclusion we have solved GLE in the limit of $\kappa \to \infty$
for the thin square film. We have predicted a series of the first and
second order phase transitions with the change of temperature and
magnetic field. On the basis of the symmetry we constructed Landau
functional for the second order transitions. First order transitions
are described on the basis of variational estimates.

We wish to thank D. Mihailovi\' c, V.V. Moshchalkov and A.S.Alexandrov
for many useful discussions and suggestions. We are also grateful to
I. Sega for critically reading the manuscript.


\begin{references}
\bibitem{mosch} L.F. Chibotaru, A. Ceulemans, V. Bruyndoncx, V.V.
Moshchalkov Nature ${\bf 408}$, 833 (2000).

\bibitem{kus}  M.B. Sobnack, F.V. Kusmartsev Phys. Rev. Lett. {\bf 86}, 716
(2001).

\bibitem{abr} A.A. Abrikosov, ZhETF, {\bf 32}, 1442 (1957).

\bibitem{bean} P.C. Bean, J.B. Livingston, Phys. Rev. Lett., {\bf 12},
14 (1964).

\bibitem{buz} A.I. Buzdin, J.P. Brison, Phys. Lett. A, {\bf 196}, 267
(1994).

\bibitem{Deo} P.S. Deo, V.A. Schweigert, F.M. Peeters, and A.K. Geim, Phys. Rev.
Lett. {\bf 79}, 4653 (1997).

\bibitem{Schw} V.A. Schweigert, F.M. Peeters, P.S. Deo,
Phys. Rev.  Lett. {\bf 81}, 2783 (1998).

\bibitem{Peet} V.A. Schweigert, F.M. Peeters, Phys. Rev. Lett. {\bf
83}, 2409 (1999).

\bibitem{km} Discrete nonlinear Schr\" odinger equation (4) is similar
to the equation describing Holstein polaron on a lattice V.V. Kabanov
O.Y.  Mashtakov, Phys. Rev. B, {\bf 47}, 6060 (1993).

\bibitem{halat} I.M. Khalatnikov 'Introduction to the theory of
superfluidity' Nauka, Moscow, 1965; W.F. Vinen, in Superconductivity, Ed. by 
R.D. Parks (Marcel Dekker, New York, 1969), Vol.2, p.1167.
\end{references}
\end{document}